\documentclass{webofc}
\usepackage[varg]{txfonts}   % Web of Conferences font
\usepackage{hyperref}
\usepackage{url}
\hypersetup{colorlinks=true,citecolor=blue,urlcolor=blue,linkcolor=blue}
\begin{document}
\title{Measurements of $|V_{cd(s)}|$ , decay constants, and form factors in (semi-)leptonic $D_{(s)}$ decays at BESIII  }

\author{\firstname{Medina} \lastname{Ablikim}\inst{1}\fnsep\thanks {\email{mablikim@ihep.ac.cn}}(on behalf of BESIII Collaboration)
        % etc.
}

\institute{Institute of High Energy Physics, Chinese Academy of Sciences, 19B Yuquan Road, Beijing 100049, China}

\abstract{The BESIII experiment accumulated the world’s largest charm-meson samples. This proceeding presents recent measurements of decay constants, transition form factors, and CKM matrix elements from (semi-)leptonic \(D^+\) and \(D_s^+\) decays based on these datasets.
}
\maketitle
\section{Introduction}
\label{intro}
Precision measurements of leptonic and semileptonic charm decays are essential for constraining the CKM matrix, which governs quark flavor and generation mixing under the weak interaction. These measurements allow extraction of decay constants and form factors parameterizing strong interaction effects. Charm physics provides fundamental information about non-perturbative QCD from decay constant and form factor measurements, and enables tests of the Standard Model (SM) via Cabibbo-Kobayashi-Maskawa (CKM) unitarity and lepton universality studies~\cite{charm1,charm2,charm3}.

The BESIII detector~\cite{bes3} has accumulated \(e^+e^-\) collision data with integrated luminosities of 20.3 fb\(^{-1}\) at \(\sqrt{s}=3.773\) GeV and 7.33 fb\(^{-1}\) in the \(\sqrt{s}=4.128\)–4.226 GeV energy region. At BESIII, experimental studies of various decays of \(D^0\), \(D^+\), \(D_s^+\) and their excited states are reviewed in Ref.~\cite{ma}, while a comprehensive overview of the amplitude-analysis methodology for charm meson decays is presented in Ref.~\cite{dong}. This proceeding mainly highlights the latest measurements of leptonic and semileptonic \(D^{0(+)}\) and \(D_s^+\) decays obtained using these two datasets. Such studies enable precision determinations of the CKM matrix elements \(|V_{cd}|\) and \(|V_{cs}|\), the decay constants \(f_{D^+}\) and \(f_{D_s^+}\), and transition form factors.

Benefiting from the clean experimental environment and low background levels inherent to \(e^+e^-\) annihilation data near threshold, the double-tag (DT) method~\cite{DT} is employed. Undetectable neutrinos are inferred using the variables \(U_{\rm miss} = E_{\rm miss} - |\vec{p}_{\rm miss}|\) and \(M^2_{\rm miss} = E_{\rm miss}^2 - |\vec{p}_{\rm miss}|^2\) to identify (semi-)leptonic signals. The branching fraction is calculated via
\begin{equation}
\mathcal{B} = \frac{N_{\rm DT}}{N_{\rm ST}\,\epsilon_{\rm DT}/\epsilon_{\rm ST}}
\end{equation}
where systematic uncertainties associated with the single-tag (ST) selection largely cancel out.
\section{Leptonic decays}
%\label{sec-1}
%
\subsection{Decay of $D^+\to \mu^+ \nu_{\mu}$ }
With the 20.3 fb\(^{-1}\) data collected at \(\sqrt{s}=3.773\) GeV, the BESIII Collaboration~\cite{Dtomunu-1} measured the branching fraction (BF) of $D^+\to \mu^+ \nu_{\mu}$ to be $(4.034 \pm 0.08_{\rm stat} \pm 0.04_{\rm syst})\times 10^{-4}$. Interpreting this measurement with knowledge of the Fermi coupling constant $G_F$, the masses of $D^+$ and $\mu^+$ as well as the life time of the  $D^+$ meson, $\tau_{D^+}$,~\cite{pdg}, the product of the decay constant and the CKM matrix element is determined to be
\begin{equation}
f_{D^+}|V_{cd}| = (48.02 \pm 0.48_{\rm stat} \pm 0.24_{\rm syst} \pm 0.12_{\rm input} \pm 0.15_{\rm EM})~\rm{MeV},
\end{equation}
where the third uncertainty arises from the input value of $\tau_{D^+}$, and the fourth is the uncertainty from the electromagnetic correction, which depends on the meson structure~~\cite{EM}. Using the value of the magnitude of the CKM matrix
element $V_{cd}$ given by the global standard model fit\cite{pdg}, the $D^+$ decay constant is extracted
\begin{equation}
f_{D^+} = (213.5\pm 2.1_{\rm stat}\pm 1.1_{\rm syst} \pm 0.8_{\rm input} \pm 0.7_{\rm EM})~\rm{MeV}.
\end{equation}
Alternatively, taking the value of $f_{D^+}$ from a precise LQCD calculation~\cite{flag}, the CKM matrix element is extracted 
\begin{equation}
|V_{cd}| = 0.2265 \pm 0.0023_{\rm stat} \pm 0.0011_{\rm syst} \pm 0.0009_{\rm input} \pm 0.0007_{\rm EM}.
\end{equation}
These are the most precise measurements. The precision is improved by a factor of 2.3 compared to the previous best measurement~\cite{Dtomunu-2}. 

\subsection{Decay of $D_s^+\to \mu^+ \nu_{\mu}$ }
Using the data with the integrated luminosity of 7.33 $\mathrm{fb}^{-1}$ collected at center-of-mass energies between 4.128 and 4.226 GeV, the BESIII Collaboration measured the BF of the leptonic decay for $D_s^+\to \mu^+ \nu_{\mu}$~\cite{Dsmunu}
\begin{equation}
\mathcal{B}( D_s^+\to \mu^+ \nu_{\mu}) = (0.5294 \pm 0.0108_{\rm stat} \pm 0.0085_{\rm syst})\%.
\end{equation}
Combining the measured BF with the world average values of $G_F$, $m_{\mu}$, $m_{D_s^+}$and the $D_s^+$ life time\cite{pdg}, the product of the $D_s^+$
decay constant $f_{D_s^+}$ and the magnitude of the $c\to s$ quark mixing matrix
element $V_{cs}$ is determined
\begin{equation}
f_{D_s^+} |V_{cs}| = 241.8 \pm 2.5_{\rm stat} \pm 2.2_{\rm syst}~\rm MeV. 
\end{equation}
Using the value of $V_{cs}$ given by the global SM fit~\cite{pdg} or the
averaged decay constant from recent LQCD calculation~\cite{lqcd1,lqcd2} as input, the decay constant $f_{D_s^+}$ and the CKM matrix element $|V_{cs}|$ are extracted to be
$f_{D_s^+} = 248.4 \pm 2.5_{\rm stat} \pm 2.2_{\rm syst}~\rm MeV $ and $|V_{cs}| = 0.968 \pm 0.010_{\rm stat} \pm 0.009_{\rm syst}$.

\begin{figure*}
\centering
\vspace*{1cm}       % Give the correct figure height in cm
\includegraphics[width=6cm,clip]{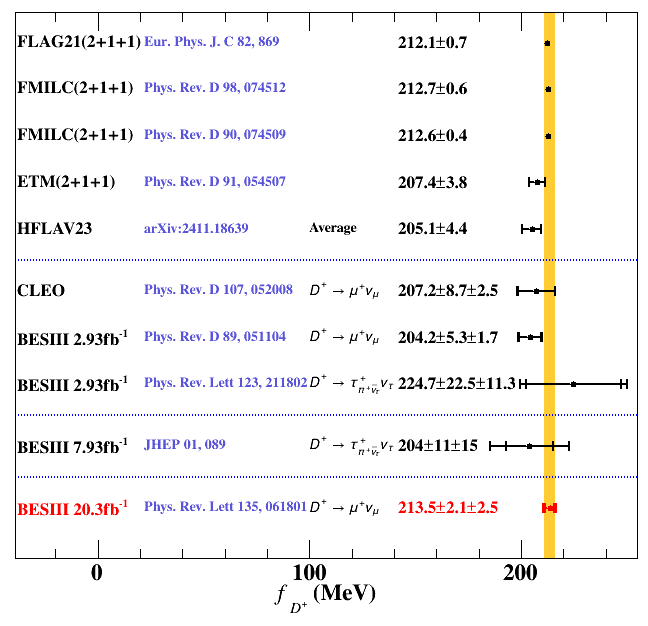}
\includegraphics[width=6cm,clip]{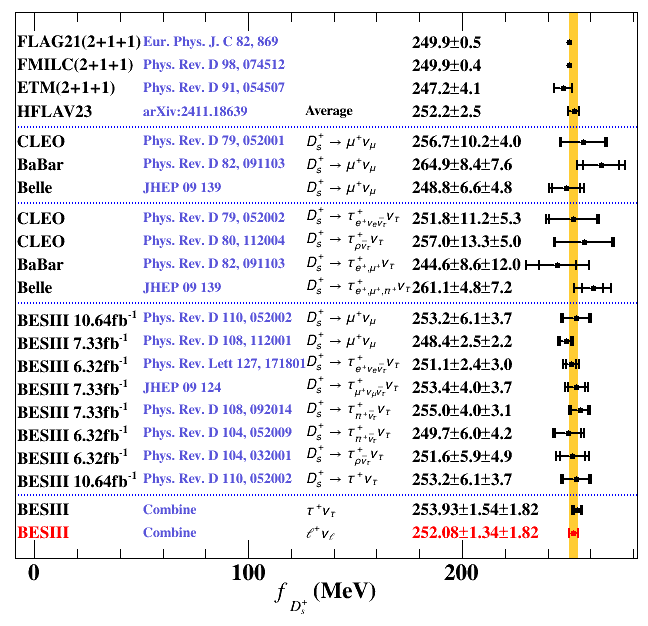}
\caption{Comparisons of the decay constants $f_{D^+}$ (left) and $f_{D_s^+}$ (right) to various experimental measurements and theoretical determinations}
\label{fig-1}       % Give a unique label
\end{figure*}

\begin{figure*}
\centering
\vspace*{1cm}       % Give the correct figure height in cm
\includegraphics[width=6cm,clip]{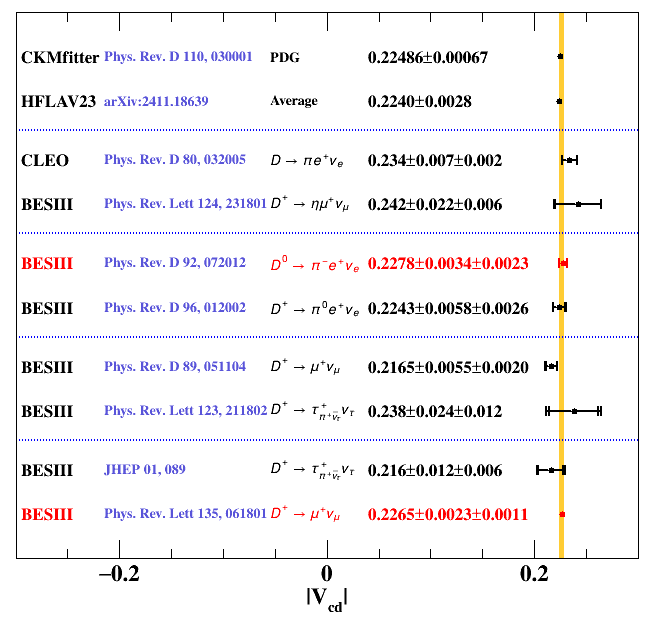}
\includegraphics[width=6cm,clip]{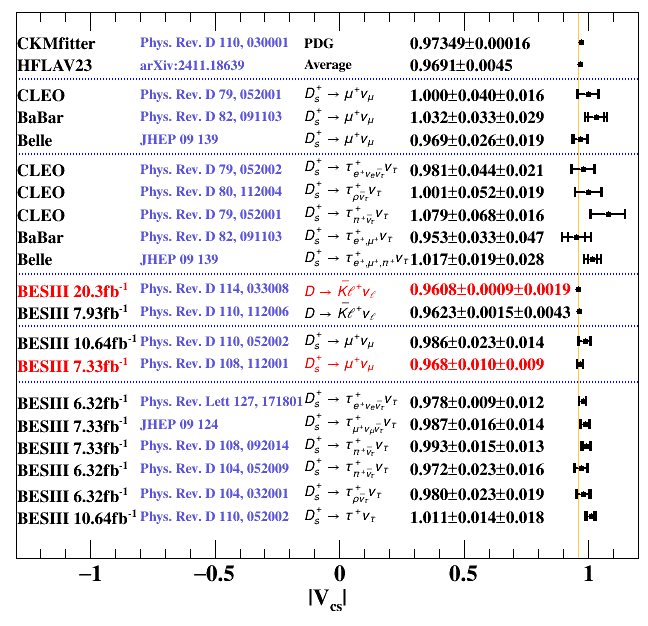}
\caption{Comparisons of the CKM matrix elements $|V_{cd}|$ (left) and $|V_{cs}|$ (right) to various experimental measurements and theoretical determinations}
\label{fig-2}       % Give a unique label
\end{figure*}

Figure~\ref{fig-1} and Figure~\ref{fig-2} summarize, respectively, the  experimental measurements and theoretical predictions for the decay constants $f_{D^+}$ and $f_{D_s^+}$, and the CKM matrix elements $|V_{cd}|$ and $|V_{cs}|$. The BESIII measurements yield the most precise determinations to date.

\section{Semileptonic Decays}
\subsection{Measurements for $D^{0(+)} \to P^{-(0)} \ell^+ \nu_\ell$ }
With the integrated luminosity of 20.3 fb\(^{-1}\) $D$ data sample, the BESIII experiment performed the precise measurements of the semileptonic decays~\cite{DtoKlnu} $D^0 \to K^- e^+\nu_e$, $D^0 \to K^- \mu^+\nu_\mu$, $D^+ \to \bar{K^0} e^+ \nu_e$ and $D^+ \to \bar{K^0} \mu^+ \nu_\mu$, and the absolute BFs measured to be $(3.548 \pm 0.006_{\rm stat} \pm 0.017_{\rm syst})\%$ , $(3.445 \pm 0.007_{\rm stat} \pm  0.017_{\rm syst})\%$, $(8.928 \pm 0.025_{\rm stat} \pm 0.050_{\rm syst})\%$,
and $(8.770 \pm 0.029_{\rm stat} \pm 0.053_{\rm syst})\%$, respectively. The
ratios of the BFs between semimuonic and semielectronic $D$ decays are determined to be 
$R^{D^0}_{\mu/e} = 0.971 \pm 0.003_{\rm{stat}} \pm 0.004_{\rm{syst}}$ and $R^{D^+}_{\mu/e} = 0.982 \pm 0.004_{\rm{stat}} \pm 0.002_{\rm{syst}}$ which exhibit a slight difference with the LFU prediction of
$0.975 \pm  0.001$~\cite{lepUniv} up to 1.5$\sigma$.  The
first measurement of the forward-backward asymmetries as well as the precise measurements of the partial decay rates of $D^{0(+)} \to \bar{K} l^+\nu_l (l = \mu,e)$ are reported~\cite{DtoKenu}. A simultaneous fit to these kinematic variables gives the first experimental constraint on the
scalar current contribution in the $D^{0(+)} \to \bar{K} l^+\nu_l$ transition. The complex Wilson coefficient is determined to be 
$Re(c_s^{\mu}) = 0.017 \pm 0.008_{\rm stat} \pm 0.006_{\rm syst}$ and $Im(c_S^\mu) = 0.077 \pm 0.011_{\rm stat} \pm 0.009_{\rm syst}$, which deviates from the SM by $2.3\sigma$. Besides, the improved measurements of the BFs and tests of the lepton-flavor universality (LFU), the hadronic transition FFs $f_+(0)$, as well as the CKM matrix element $|Vcs|$, which are in the highest precision to date, were also presented. 

Using the same dataset, the first observation of the semileptonic decay $D^+ \to \eta' \mu^+ \nu_\mu$ with a significance of $8.6\,\sigma$ and an improved measurement of $D^+ \to  \eta' e^+ \nu_e$ are reported~\cite{Dtoetapenu}. The BFs of $D^+ \to  \eta' \mu^+ \nu_\mu$ and $D^+ \to  \eta' e^+ \nu_e$ are determined to be $(1.92 \pm 0.28_{\mathrm{stat}} \pm 0.08_{\mathrm{syst}}) \times 10^{-4}$ and $(1.79 \pm 0.19_{\mathrm{stat}} \pm 0.07_{\mathrm{syst}}) \times 10^{-4}$, respectively. The ratio of the two BFs is determined to be $R_{\mu/e} = 1.07 \pm 0.19_{\mathrm{stat}} \pm 0.03_{\mathrm{syst}}$, which agrees with the theoretical expectation of lepton flavor universality within the SM. From an analysis of the $D^+ \to  \eta' \ell^+ \nu_l$ decay dynamics, the product of the hadronic form factor (FF) $f_+^{\eta'}(0)$ and the CKM matrix element $|V_{cd}|$ is measured for the first time, giving $f_+^{\eta'}(0)\,|V_{cd}| = (5.92 \pm 0.56_{\mathrm{stat}} \pm 0.13_{\mathrm{syst}}) \times 10^{-2}$. Combining with the BFs of $D^+ \to \eta \ell^+ \nu_\ell$~\cite{pdg} and $D_s^+ \to \eta^{(\prime)} \ell^+ \nu_\ell$~\cite{Dstoetapenu,Dstoetaenu}, the $\eta$-$\eta'$ mixing angle in the quark flavor basis is determined to be $\phi_P = (39.8 \pm 0.8_{\mathrm{stat}} \pm 0.3_{\mathrm{syst}})^\circ$, which provides complementary
data to constrain the gluon component in the $\eta'$ meson.

\subsection{Measurements for $D^+ \to V \ell^{+} \nu_\ell$ }

The BESIII experiment reported the first measurement of full set of $CP$ asymmetries and averaged angular observables with $A_{CP}$ in the decays $D^+ \to \bar{K}^{*0} \ell^+ \nu_\ell$  via $\bar{K}^{*0} \to K_S^0\pi^0$~\cite{DtoVenu}. In addition, the absolute BF of $D^+ \to K_S^0 \pi^0 \mu^+ \nu_\mu$ is measured to be $(0.943 \pm 0.012_{\rm stat} \pm 0.010_{\rm syst}) \%$ for the first time, and the absolute BF of $D^+ \to K_S^0 \pi^0 e^+ \nu_e$ is measured with improved precision. By analyzing the $D^+ \to K_S^0 \pi^0 \ell^+ \nu_\ell$ decay dynamics, the fraction of the P-wave component and the hadronic FF ratios of $D^+ \to \bar{K}^{*0}$ were presented. The LFU was tested, and no evidence for violation of $\mu-e$ LFU was observed in either the full range or the five chosen bins of momentum-transfer squared.

\subsection{Measurements for $D_s^+ \to S \ell^{+} \nu_\ell$ }

Based on the $D_s$ data sample of $7.33\,\mathrm{fb}^{-1}$, an analysis of the decay $D_s^+ \to \pi^+\pi^-e^+\nu_e$ via the intermediate resonant state $f_0(980) \to \pi^+\pi^-$ has been performed~\cite{DstoSenu}. The BF of the  $D_s^+ \to f_0(980)e^+\nu_e$ decay with the subsequent decay $f_0(980) \to \pi^+\pi^-$ is measured to be $(1.72 \pm 0.13 \pm 0.10)\times 10^{-3}$. By investigating the decay dynamics with a simple pole parameterization for the FF and the Flatté formula to characterize the $f_0(980)$ contribution in the differential decay rate, the product of the hadronic FF $f_+^{D_s\to f_0(980)}(0)$ and the $c\to s$ CKM matrix element $|V_{cs}|$ is determined for the first time, yielding $f_+^{D_s\to f_0(980)}(0)|V_{cs}| = 0.504 \pm 0.017_{\rm stat} \pm 0.035_{\rm syst}$. Furthermore, the decay $D_s^+ \to f_0(500)e^+\nu_e$ is searched for the first time but no signal is found. The upper limit on the BF of $D_s^+ \to f_0(500)e^+\nu_e$, $f_0(500) \to \pi^+\pi^-$ decay is set to be $3.3 \times 10^{-4}$ at 90\% confidence level.

\subsection{Measurements for $D^+_s \to A \ell^{+} \nu_\ell$ }

Using  $7.33\,\mathrm{fb}^{-1}$ of data at  $4.128$–$4.226$ GeV, the semileptonic decays
$D^+_s \to K_1(1270)^0 e^+\nu_e$ and $D^+_s \to b_1(1235)^0 e^+ \nu_e$~\cite{DstoAenu1} as well as
$D^+_s \to f_1(1420) e^+ \nu_e$ and $D^+_s \to f_1(1285) e^+ \nu_e$~\cite{DstoAenu2} are also 
searched for the first time. The first evidence for the decay $D^+_s \to f_1(1420) e^+ \nu_e$ is found with a statistical significance of $3.4\,\sigma$, and its product branching fraction 
$\mathcal{B}[D^+_s \to f_1(1420) e^+\nu_e]$· $\mathcal{B}[f_1(1420) \to K^+ K^- \pi^0]$ is determined to be $(4.5^{+2.0}_{-1.7}) \times 10^{-4}$. No significant signals of $D^+_s \to K_1(1270)^0 e^+\nu_e$ and $D^+_s \to b_1(1235)^0 e^+ \nu$, and $D^+_s \to f_1(1285) e^+\nu_e$ are observed and the upper limit on their product BF at the 90\% confidence level are set to be $\mathcal{B}[D^+_s \to K_1(1270)^0 e^+\nu_e] < 4.1 \times 10^{-4}$, $\mathcal{B}[D^+_s \to b_1(1235)^0 e^+\nu_e]$·$\mathcal{B}[b_1(1235)^0 \to \omega \pi^0] < 6.4 \times 10^{-4}$, and $\mathcal{B}[D^+s \to f_1(1285) e^+\nu_e]$·$\mathcal{B}[f_1(1285) \to \pi^+\pi^-\eta] < 1.7 \times 10^{-4}$, respectively.

\section{Summary and outlook}

The BESIII experiment has achieved world-leading precision measurements for leptonic and semileptonic decays of $D^+$ and $D_s^+$: the decay constants are measured to a precision of about 1\%, and the CKM matrix elements $|V_{cs}|$ and $|V_{cd}|$ are extracted to precisions of 0.5\% and 1\%, respectively. Looking ahead, many new and updated measurements are expected using the BESIII charm datasets.\cite{futureProg}.

% BibTeX or Biber users please use (the style is already called in the class, ensure that the "woc.bst" style is in your local directory)
% \bibliography{your_bib_file} % Replace "your_bib_file" with the actual name of your .bib file

\end{document}